\documentclass[10pt,conference]{IEEEtran}
\usepackage{graphicx} 
\usepackage{amsmath} 
\usepackage{enumitem}
\usepackage{booktabs}
\usepackage{tabularx}
\usepackage{cite}
\usepackage{multirow}
\usepackage{amssymb}
\usepackage{url}
\usepackage{subcaption}
\usepackage{stfloats}
\usepackage[T1]{fontenc}

\title{Fidelity-Guaranteed Entanglement Routing with Distributed Purification Planning}
\author{
  \IEEEauthorblockN{Anthony Gatti, Anoosha Fayyaz, Prashant Krishnamurthy, Kaushik P. Seshadreesan, and Amy Babay}
  \IEEEauthorblockA{University of Pittsburgh}
  \IEEEauthorblockA{\{amg671, anf224, prashk, kausesh, babay\}@pitt.edu}
}

\begin{document}

\maketitle

\begin{abstract}
    Many quantum-network applications require end-to-end Bell pairs whose fidelity exceeds a request-specific threshold, but existing entanglement routing algorithms either optimize only throughput without regard for fidelity or enforce fidelity guarantees using centralized controllers with global link-state knowledge. We present Q-GUARD, an online entanglement routing algorithm that enforces per-request fidelity thresholds within a distributed protocol model in which nodes exchange link-state information only with their $k$-hop neighbors. After link outcomes are realized in each slot, Q-GUARD builds per-link purification cost tables from realized Bell pairs, allocates per-hop fidelity targets using a Werner-state equal-split rule, and selects between candidate path segments using a segment-local expected-goodput (EXG) metric that jointly accounts for swap success, purification overhead, and resource availability. We also introduce Q-GUARD-WS, an extension that exploits per-link hardware quality estimates to allocate purification effort non-uniformly across hops. On synthetic 100-node topologies with heterogeneous link fidelity and stochastic BBPSSW purification, Q-GUARD raises the qualified success rate from under 20\% to over 85\% on 4-hop paths and nearly doubles the qualified service radius in Euclidean distance relative to throughput-only and naive-purification baselines, while Q-GUARD-WS provides additional throughput gains under high hardware heterogeneity.
\end{abstract}

\section{Introduction}
\label{sec:introduction}

Quantum networks distribute entanglement between remote nodes to enable applications such as quantum key distribution (QKD), distributed quantum computing, and quantum sensing \cite{Kimble08,WehnerRoadmap,RFC9340}. Long-distance entanglement is created by joining short-range Bell pairs through entanglement swapping at intermediate repeaters \cite{Zukowski93Swapping, Briegel98}. The process of determining how to create and combine short-range entanglements to produce long-distance end-to-end entanglement is known as \emph{entanglement routing}. Every swap operation and every imperfect link degrades the fidelity of the resulting end-to-end entangled state. This degradation presents severe issues in many applications: in QKD, the quantum bit-error rate grows with decreasing fidelity until the secret key rate drops to zero \cite{Ekert91}; in distributed quantum computing, gate teleportation with low-fidelity ebits introduces logical errors that compound across circuit depth \cite{RFC9583}; and in entanglement-enhanced sensing, sub-threshold fidelity causes precision to degrade toward the classical shot-noise limit \cite{WehnerRoadmap}. Many applications therefore require \emph{fidelity-qualified} ebits: end-to-end Bell pairs whose fidelity exceeds a request-specific threshold $F_{\mathrm{th}}$. 

Fidelity-aware entanglement routing protocols have begun to be introduced to address this need. EFiRAP \cite{EFiRAP} jointly optimizes routing and purification to maximize the number of connections meeting a fidelity threshold. Q-LEAP \cite{QLEAP} introduces a less computationally intensive approach using an equal-split heuristic that assigns each hop a uniform fidelity target to all links. However, these designs assume routing swapping, and purification decisions are made by a centralized controller with global link-state knowledge. This assumption limits their applicability to realistic deployments, where link-state information must be exchanged over classical channels with non-negligible latency at network scale and entangled pairs decohere in quantum memory while awaiting routing decisions.

Decoherence makes \emph{distributed} operation particularly important for fidelity-guaranteed routing. In any entanglement routing protocol, generated Bell pairs must be held in quantum memory while link-state information is collected and routing decisions are made. During this time, the pairs decohere, so the fidelity at the time of use is lower than the fidelity at the time of generation. In a centralized protocol, link-state information must traverse the entire network to reach the controller, and routing decisions must be disseminated back, introducing communication delays that scale with network diameter. A distributed protocol where nodes make decisions using only \emph{$k$-hop local link-state information} incurs significantly less communication overhead: each node only exchanges entanglement generation outcomes with neighbors up to $k$ hops away, and decisions can be made as soon as this local exchange completes. The result is that entangled pairs spend less time in memory before being consumed by purification and swapping operations, reducing the gap between initial and time-of-use fidelity. 

Q-CAST~\cite{QCAST} provides a natural foundation for \emph{distributed} fidelity-aware routing. It introduced a comprehensive model for decentralized entanglement routing with concurrent source--destination pairs, limited qubit capacity, and $k$-hop local link-state exchange, organizing each time slot into phases for request dissemination, path selection, link generation, and recovery. Q-CAST achieves high throughput, but optimizes only for rate: it delivers pairs regardless of fidelity, without purification planning or per-request threshold enforcement. No existing algorithm provides per-request fidelity enforcement within this locality-constrained, distributed framework. 

The need for such algorithms is becoming increasingly urgent. Metropolitan-scale quantum networks are advancing from two-node demonstrations to multi-node testbeds~\cite{pompili2021realization, Liu24, stolk2024metropolitan}, with recent work demonstrating memory--memory entanglement across metropolitan distances on deployed fiber~\cite{Liu24, stolk2024metropolitan} and entanglement swapping on deployed urban fiber~\cite{bersin2024development, craddock2026high}. As these testbeds scale to tens or hundreds of nodes, fidelity-aware routing that operates within realistic locality constraints will be essential.

We present \textbf{Q-GUARD} (Guaranteed-fidelity Adaptive Routing with Detours), a distributed entanglement routing algorithm that bridges this gap. Q-GUARD preserves Q-CAST's distributed, $k$-hop protocol structure but adds a fidelity-aware planning phase that runs after link outcomes are known in each slot. Nodes assign per-hop fidelity targets derived from the requested threshold $F_{\mathrm{th}}$, build per-link purification cost tables from the realized Bell pairs, and use a segment-local \emph{expected goodput} (EXG) metric that incorporates fidelity to choose between candidate path segments and recovery detours. We also introduce \textbf{Q-GUARD-WS} (Weighted Split), an extension that can allocate purification effort non-uniformly across hops to account for hardware heterogeneity, and evaluate \textbf{Q-GUARD-FP} (Fidelity-aware Planning), an exploratory variant that incorporates purification cost into its path finding.

Our primary contributions are:
\begin{itemize}[leftmargin=1.5em, itemsep=2pt, topsep=3pt]
    \item We design a fidelity-aware routing algorithm that enforces per-request end-to-end fidelity thresholds within a distributed, $k$-hop local link-state model. Q-GUARD adds per-hop purification planning and segment-local EXG-based recovery selection without requiring global state or additional controller messages.
    \item We introduce Q-GUARD-WS, a weighted-split extension that uses predicted initial link fidelities to allocate purification effort non-uniformly across hops, reducing unnecessary purification on higher quality links and improving service in networks with heterogeneous hardware.
    \item We evaluate Q-GUARD and Q-GUARD-WS against Q-CAST and Q-CAST-PUR (Q-CAST with end-to-end purification) on synthetic topologies with heterogeneous link fidelity and stochastic BBPSSW purification\cite{Bennett96Purification}. Q-GUARD extends the effective range of fidelity-qualified service in both hop count and Euclidean distance, and consistently delivers higher qualified throughput across a range of operating conditions. Q-GUARD-WS provides further improvements as hardware heterogeneity increases.
\end{itemize}

\section{Related Work}
\label{sec:related-work}

Entanglement routing has attracted substantial attention across multiple design dimensions; Abane et al.~\cite{Abane25} provide a comprehensive survey, noting among other findings that per-link fidelity estimation remains non-trivial in deployed networks, posing a challenge that Q-GUARD addresses through its purification planning methodology. We focus here on the protocol categories most relevant to Q-GUARD's contributions.

\paragraph{Fidelity-guaranteed routing with global state}
EFiRAP~\cite{EFiRAP} was the first to quantify end-to-end fidelity under bit-flip errors and to identify critical links for resource-efficient purification, jointly optimizing routing and purification via LP-based path selection. Q-PATH and Q-LEAP \cite{QLEAP} target the same objective: Q-PATH finds the minimum-cost path and purification plan using iterative search, while Q-LEAP reduces computational complexity with an equal-split rule that assigns each hop a uniform fidelity target $F_{\mathrm{avg}} = F_{\mathrm{th}}^{1/L}$. Q-GUARD adopts the Q-LEAP equal-split rule and extends it with a non-uniform, hardware-weighted variant (Q-GUARD-WS). PERA \cite{PERA} provides hop-by-hop fidelity guarantees using multiple purification attempts within the decoherence window. All of these designs assume a centralized controller with global link-state knowledge and plan routing/purification using this global information. None operate within a distributed, $k$-hop local link-state model.

\paragraph{Throughput-oriented routing with local state}
Q-PASS and Q-CAST \cite{QCAST} introduced the time-slotted, $k$-hop local link-state model for entanglement routing with multiple concurrent flows, limited qubit capacity, and contention-free path selection via a non-additive expected throughput metric (EXT). Q-CAST optimizes only for for overall end-to-end entanglement rate (throughput), however, and uses no fidelity modeling or purification. Centralized throughput-oriented designs such as MULTI-R \cite{MULTIR} similarly lack per-request fidelity guarantees. Caleffi \cite{Caleffi17} established that quantum routing metrics are fundamentally non-additive, which is true of both Q-CAST's EXT and our EXG metric.

\paragraph{Multi-path and rate--distance routing}
Like Q-CAST, Q-GUARD can use multiple paths to increase the end-to-end entanglement rate between a pair of nodes. Pant et al. \cite{Pant20} showed that multi-path strategies on grid topologies can achieve distance-independent entanglement rates by exploiting path diversity. Li et al. \cite{Li21} integrated multi-path selection with purification under a centralized model. Chakraborty et al. \cite{Chakraborty20} formulated entanglement distribution as a multi-commodity flow problem with LP-based rate optimization. These works demonstrate the importance of path diversity and resource allocation, but none operate under $k$-hop locality constraints or enforce fidelity requirements in a distributed setting.

\paragraph{Purification and swap scheduling}
Complementary to routing, several works optimize purification and swapping along a \emph{given} path. FMSPP \cite{FMSPP} proposes prioritized-purification scheduling with concurrent swapping. PSC \cite{PSC} addresses network-level purification resource allocation for concurrent requests. Chen and Jia \cite{PurSched} develop optimal single-hop purification schedules and extend the framework to fidelity-constrained multi-flow routing. Hu et al. \cite{Hu21} demonstrated long-distance entanglement purification experimentally, providing physical validation for purification-based models. These works provide useful building blocks, but they do not fit Q-CAST’s decentralized model: FMSPP assumes scheduling on a globally known path, PSC assumes centralized network-level purification control, and Chen and Jia formulate purification-aware routing as a global optimization problem. Q-GUARD instead targets purification-aware routing within Q-CAST’s online, k-hop local-state model.

\section{Model}
\label{sec:model}

\subsection{Timing and Link-State Communication Model}

We adopt the time-slotted, $k$-hop local link-state model introduced by Q-CAST~\cite{QCAST}. Within each slot, links attempt entanglement generation, nodes exchange outcomes within $k$ hops via classical signaling, and swapping decisions are made based on this local view. We assume that static network topology information is known in advance, and that at the start of each slot, a controller broadcasts the active request set to all nodes; otherwise, no global state or coordination is needed.

\subsection{Network and Noise Model}

We model the quantum network as an undirected graph $G = (V, E)$. Each node $v \in  V$ is a quantum switch (or end host) with a finite number of qubit memories and classical communication links with its neighbors. Each edge $e = (u, v) \in E$ represents a bidirectional quantum channel between $u$ and $v$. Each link~$e$ is characterized by a generation success probability $p_e = e^{-\alpha \ell_e}$ \cite{Pirandola17} (where $\ell_e$ is the physical length and $\alpha$ is an attenuation constant) and an initial Bell-pair fidelity $F_e^{(0)}$ for successfully generated pairs. Entanglement swapping at each intermediate node succeeds with a fixed probability $q$ that is the same for all nodes.

We model all bipartite entanglement as Werner states~\cite{Werner89, Kalb17} with fidelity $F$ and Werner parameter $w(F) = (4F - 1)/3$. Under this model, entanglement swapping is multiplicative: swapping two Werner states with parameters $w_1, w_2$ yields a Werner state with $w_{\mathrm{out}} = w_1 w_2$. The end-to-end fidelity of an $L$-hop path is therefore
\begin{equation}
  F_{\mathrm{end}} = \frac{1 + 3\prod_{i=1}^{L} w(F_i)}{4}.
  \label{eq:e2e}
\end{equation}

\paragraph{Heterogeneous initial fidelity}
In heralded entanglement generation protocols, the fidelity of a successfully generated Bell pair is determined primarily by local hardware characteristics: detector efficiency, memory quality, mode matching \cite{Dahlberg19}. To model the hardware heterogeneity present in deployments with mixed-generation  or multi-vendor equipment, we assign each link $e$ a hardware quality parameter $\eta_e$ drawn independently from a normal distribution $\mathcal{N}(\bar{\eta},\,\sigma_\eta)$, where $\sigma_\eta$ controls the degree of network heterogeneity. The initial fidelity of a freshly generated Bell pair on link $e$ is 
\begin{equation}
  F_e^{(0)} = \tfrac{1}{4} + \tfrac{3}{4}\,e^{-1/\eta_e}
  \label{eq:f0}
\end{equation}
with per-generation Gaussian noise ($\sigma = 0.01$) to model shot-to-shot variability. Higher $\eta_e$ corresponds to better hardware and higher initial fidelity. Unlike models that assume uniform $F_0$ across all links \cite{QLEAP, Chakraborty20}, this produces heterogeneous initial fidelities. This better captures realistic deployment conditions, which motivates the need for non-uniform purification allocation in Q-GUARD-WS (Section \ref{sec:ws}).

\paragraph{Purification model}
We model entanglement purification using the symmetric BBPSSW protocol for Werner states \cite{Bennett96Purification}. Each round takes two Bell pairs as input and produces one output pair with improved fidelity. Success for this operation is probabilistic: for inputs with fidelities $F_a, F_b$, the output fidelity and success probability are determined by the BBPSSW recurrence \cite{Bennett96Purification, Deutsch96Purification}. In our analytic cost tables, we use the symmetric case $F_a = F_b$ to estimate minimum round counts (Section \ref{sec:phase4}); in runtime execution, we allow asymmetric inputs by selecting the best available pair from the realized pool on each hop (Section~\ref{sec:phase5}).

\subsection{Requests and service objective}

A protocol request is a tuple $(s, d, F_{\mathrm{th}}$), where $s,d \in V$ are the source and destination and $F_{\mathrm{th}} \in (0,1)$ is the minimum fidelity requested by the higher layer. In each slot $t$, let $X_t(s,d,F_{\mathrm{th}})$ denote the number of end-to-end pairs delivered for this request with fidelity (under the Werner model, Eq.~\eqref{eq:e2e}) of at least $F_{\mathrm{th}}$. We define the \emph{fidelity-qualified goodput} over $T$ timeslots as:
\begin{equation}
  G(s,d,F_{\mathrm{th}}) = \frac{1}{T}\sum_{t=1}^{T} X_t(s,d,F_{\mathrm{th}}).
  \label{eq:goodput}
\end{equation}
Our objective is to maximize total network goodput $\sum_{(s,d,F_{\mathrm{th}})} G(s,d,F_{\mathrm{th}})$, subject to the locality and resource constraints of the $k$-hop slotted model. Lower-fidelity pairs may be created and consumed internally by the protocol (e.g., as purification inputs), but they are not counted as delivered ebits and do not contribute to total goodput.

\begin{table}[t]
\small
\centering
\caption{Summary of notation.}
\label{tab:notation}
\renewcommand{\arraystretch}{0.92}
\begin{tabularx}{\columnwidth}{@{}lX@{}}
\toprule
Symbol & Description \\
\midrule
$G = (V, E)$ & Quantum network multigraph \\
$p_e$ & Link generation success probability \\
$q$ & Swap success probability \\
$k$ & Link-state broadcast radius (hops) \\
$d$ & Average degree of each node \\
$F$, $w$ & Fidelity and Werner parameter ($w = \tfrac{4F-1}{3}$) \\
$F_{\mathrm{th}}$, $w_{\mathrm{th}}$ & Requested end-to-end fidelity threshold \\
$F_e^{(0)}$ & Initial fidelity of Bell pairs on link $e$ \\
$\eta_e$ & Hardware quality parameter for link $e$ \\
$C_{\mathrm{pur}}(e, F')$ & Min.\ purification rounds on $e$ to reach $F'$ \\
$R_{\max}$ & Maximum purification rounds per link \\
$r_e$ & Assigned purification rounds for hop $e$ \\
$W$ & Path width (min.\ parallel channels per hop) \\
$L$ & Path length in hops \\
$Q_k^i$ & Prob.\ hop $k$ yields exactly $i$ successful links \\
$P_k^i$ & Prob.\ bottleneck width across hops $1..k$ equals $i$ \\
$\mathrm{EXT}(P)$ & Expected throughput of path $P$ \\
$\mathrm{EXG}(P)$ & Expected goodput of span $P$ \\
$A_{\min}(P)$ & Bottleneck availability factor for span $P$ \\
$D_{\mathrm{seg}}$, $D_{\mathrm{total}}$ & Depolarization weights (Eq.~\ref{eq:ws-seg}) \\
$c_k^*$ & Purification cost on hop $k$: $2^{r_k}$ raw pairs per output \\
$M^*$ & Max.\ purified bottleneck width: $\min_k \lfloor W/c_k \rfloor$ \\
$Q_k^{\prime\,i\,*}$, $P_k^{\prime\,i\,*}$ & Purification-aware analogs of $Q_k^i$, $P_k^i$ \\
$G(s,d,F_{\mathrm{th}})$ & Fidelity-qualified goodput for request $(s,d,F_{\mathrm{th}})$ \\
\midrule
\multicolumn{2}{@{}l}{\footnotesize $^*$Used only in Q-GUARD-FP (Section~\ref{sec:fp}).} \\
\end{tabularx}
\end{table}

\section{Algorithm: Q-GUARD}
\label{sec:algorithm}

We now present Q-GUARD, a fidelity-aware entanglement routing algorithm that augments Q-CAST’s time-slotted, $k$-hop local link-state protocol with per-hop purification planning and fidelity-qualified span selection. All decisions use only the $k$-hop state already exchanged in Q-CAST's protocol; no additional controller messages are introduced. Q-GUARD's additional computation is bounded by the link-state broadcast radius $k$, the maximum node degree, and the maximum number of purification rounds. Since these are all small constants in practice, our fidelity-aware planning is a relatively lightweight addition. A detailed complexity analysis is provided in the appendix. 

\subsection{Overview}
\begin{figure}[t]
  \centering
  \includegraphics[width=\columnwidth]{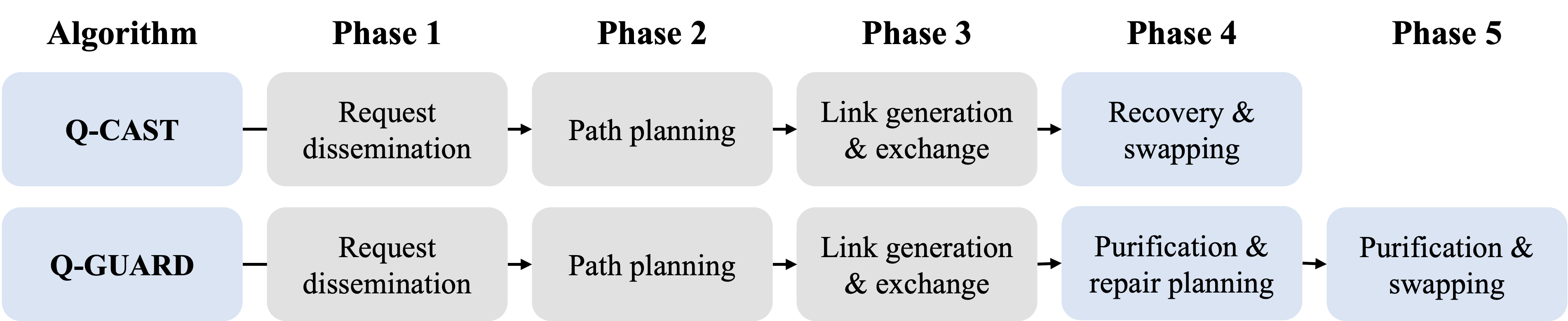}
  \caption{Q-CAST vs.\ Q-GUARD routing pipelines. Phases 1--3 are shared. Q-GUARD diverges in Phase 4, adding fidelity-aware recovery planning with per-hop purification targets and an EXG-based segment selection metric. Phase 5 executes purification, swapping, and fidelity qualification.}
  \label{fig:pipeline}
\end{figure}

Q-GUARD organizes each time slot into five phases illustrated in Figure~\ref{fig:pipeline}. Phases 1--3 are identical to Q-CAST \cite{QCAST}; the controller disseminates active requests (Phase 1), nodes select contention-free paths and reserve resources (Phase 2), and links attempt entanglement generation followed by $k$-hop link-state exchange (Phase 3). Q-GUARD diverges in Phase 4, where it adds fidelity-aware recovery planning: nodes compute per-hop fidelity targets from the requested threshold $F_{\mathrm{th}}$, build purification cost tables from the realized Bell pairs, and use a segment-local \emph{expected goodput} (EXG) metric to choose between candidate path segments. Finally, Phase 5 executes purification and entanglement swapping along the chosen spans, and only end-to-end Bell pairs meeting $F_{\mathrm{th}}$ are delivered and counted towards throughput. The following subsections describe each phase in detail. Sections~\ref{sec:ws} and~\ref{sec:fp} then present Q-GUARD-WS and Q-GUARD-FP, two variants that exploit hardware quality estimates to modify purification allocation and path scoring, respectively.

\subsubsection{Phase 1: Request Dissemination}
\label{sec:phase1}
At the start of each time slot, a controller disseminates the set of active source--destination requests to all nodes. The network topology $G = (V, E)$, along with per-node qubit capacities, the number of parallel channels on each edge, and physical link lengths (which determine generation success probabilities $p_e$), is assumed to be relatively stable and known to all nodes in advance (e.g., from an initial configuration phase). Only the per-slot request set is broadcast in each time slot. This phase is identical to Q-CAST's Phase 1, with the addition that Q-GUARD variants using $\eta_e$ (Sections~\ref{sec:ws} and~\ref{sec:fp}) require these parameters to be included in the static topology.

\subsubsection{Phase 2: Path Selection and Resource Reservation}
\label{sec:phase2}
In Phase 2, nodes select paths for the active requests, and reserve qubit and channel resources along those paths. All run the same deterministic algorithm on the same inputs (topology and request set), so the selected paths are globally consistent without additional communication. This phase is identical to Q-CAST's Phase 2; we describe it here because the concepts of major paths, recovery paths, width, and the EXT metric are essential to understanding Q-GUARD's fidelity-aware extensions in Phase 4. 

\paragraph{Path width}
A \emph{$(W, L)$-path} is a path of $L$ hops in which every hop has at least $W$ parallel channels available. Each hop corresponds to a link between two adjacent nodes. When a hop is selected as part of a path, each of the $W$ channels reserves one memory qubit at both endpoint nodes of that link, so that entanglement generation can be attempted on all $W$ channels in Phase 3. Width is important because a $W$-wide path can tolerate link failures more robustly than $W$ disjoint single-channel paths using the same resources: it fails only when all $W$ channels on a single hop fail simultaneously~\cite{QCAST}.

\paragraph{Expected throughput (EXT)}
To evaluate candidate paths, Q-CAST defines the \emph{expected throughput} (EXT) metric, which computes the expected number of end-to-end ebits a $(W, L)$-path will deliver in a single time slot. Let $p_k$ denote the channel success probability on hop~$k$. Define $Q_k^i$ as the probability that hop~$k$ produces exactly $i$ successful links:
\begin{equation}
  Q_k^i = \binom{W}{i}\, p_k^{\,i}\,(1 - p_k)^{W - i}.
  \label{eq:ext-q}
\end{equation}
The end-to-end throughput is limited by the \emph{bottleneck}: the hop with the fewest successful links. Let $P_k^i$ denote the probability that the minimum number of successful links across hops $1, \ldots, k$ is exactly~$i$:
\begin{align}
  P_1^i &= Q_1^i \notag \\[4pt]
  P_k^i &= P_{k-1}^i \cdot \sum_{l=i}^{W} Q_k^l
          \;+\; Q_k^i \cdot \sum_{l=i+1}^{W} P_{k-1}^l
  \label{eq:ext-bottleneck}
\end{align}
for $k = 2, \ldots, L$ and $i = 0, 1, \ldots, W$. Including the swap success probability $q$ at each of the $L{-}1$ intermediate nodes, the expected throughput is
\begin{equation}
  \mathrm{EXT}(P) = q^{\,L-1} \cdot \sum_{i=1}^{W} i \cdot P_L^i.
  \label{eq:ext}
\end{equation}
EXT is non-additive, i.e. it cannot be decomposed as a sum of per-hop costs, but decreases monotonically as a path is extended, since adding a hop can only reduce the bottleneck width and introduces an additional swap factor~$q$. This monotonicity property allows EXT to be used as the objective in an Extended Dijkstra Algorithm (EDA) that finds the path with maximum EXT between any source--destination pair~\cite{QCAST}.

\paragraph{Major path selection}
Paths are selected greedily to avoid resource contention. For each active request, EDA finds the path with the highest EXT in the current residual graph (the topology with previously reserved resources removed). Among all requests, the path with the highest EXT is selected and its resources (qubits and channels on every hop) are exclusively reserved. The residual graph is updated, and the process repeats until no further paths can be found. The resulting paths are called \emph{major paths}; they are contention-free by construction, meaning the network can simultaneously satisfy the resource requirements of all selected major paths.

\paragraph{Recovery path selection}
After major paths are reserved, remaining resources in the residual graph are used to select \emph{recovery paths}. A recovery path is a short alternative route whose two endpoints both lie on a single major path, with the endpoints separated by at most $k$ hops along the major path. This constraint ensures that in Phase 3, when nodes exchange link-state information within $k$ hops, all nodes on a recovery path know whether the corresponding major-path segment has failed and can make consistent repair decisions. Recovery paths are also selected via EDA in the residual graph, and their resources are exclusively reserved. Each major-path segment may have multiple candidate recovery paths available.

\subsubsection{Phase 3: Link Generation and Link-State Exchange}
\label{sec:phase3}
All channels on reserved major and recovery paths attempt entanglement generation. We assume each channel succeeds independently with probability $p_e = e^{-\alpha \ell_e}$, producing a Bell pair with initial fidelity $F_e^{(0)}$ determined by the link's hardware quality and per-generation noise (Eq. \eqref{eq:f0}, Section \ref{sec:model}). Each node then exchanges link outcomes with all nodes within $k$~hops via classical signaling. In Q-CAST, this exchange communicates only success or failure of each channel. Q-GUARD extends this to also include the realized initial fidelity of each successfully generated pair, which Phase 4 requires for purification planning. This uses the same $k$-hop communication structure with a small increase in per-link information; no additional rounds of communication are introduced.

\subsubsection{Phase 4: Fidelity-Aware Recovery Planning}
\label{sec:phase4}
In Q-CAST's Phase 4, each node uses the $k$-hop link-state information from Phase 3 to repair broken major paths: if any hop has no successful link, the algorithm covers the failed segment with a recovery detour, preferring shorter detours since fewer swaps yield higher EXT. Once the final path is assembled, entanglement swapping is performed at each intermediate node.

Q-GUARD replaces this recovery logic with a fidelity-aware planning procedure. Swapping is deferred to Phase 5; Phase 4 produces a final assembled route and a per-hop purification plan. The procedure has three steps, described below.

\paragraph{Step 1: Per-hop fidelity target allocation}
Under the Werner noise model, the end-to-end fidelity of an $L$-hop path satisfies Eq. \eqref{eq:e2e}, and the threshold $F_{\mathrm{th}}$ is met when $\prod_{i=1}^{L} w(F_i) \geq w_{\mathrm{th}} = (4F_{\mathrm{th}} - 1)/3$. Q-GUARD allocates per-hop targets by applying the equal-split rule from Q-LEAP \cite{QLEAP}: each hop receives the same target Werner parameter $w_{\mathrm{avg}} = w_{\mathrm{th}}^{1/L}$, corresponding to the per-hop target fidelity
\begin{equation}
  F_{\mathrm{avg}} = \frac{1 + 3\,w_{\mathrm{th}}^{1/L}}{4}.
  \label{eq:uniform-target}
\end{equation}
This rule is well-suited to the distributed $k$-hop model: each node sees realized fidelities only within its $k$-hop neighborhood, so no single node has visibility into the realized link quality of every hop on the path. The equal-split rule lets every node independently compute the same per-hop target from only the path length and the requested threshold, with no coordination about distant hops' outcomes. Q-GUARD-WS (Section \ref{sec:ws}) relaxes this by using hardware quality parameters $\eta_e$, which are static topology information known to all nodes in advance, to allocate targets non-uniformly.

When a recovery detour of $l'$ hops replaces a segment of $l$ ($l \leq L$) hops on the major path, the detour inherits that segment's share of the total Werner budget (its allocated share of the end-to-end Werner threshold): $w_{\mathrm{seg}} = w_{\mathrm{th}}^{l/L}$. This budget is then split uniformly across the detour's $l'$ hops, giving each hop a target of $w_{\mathrm{detour}} = w_{\mathrm{seg}}^{1/l'}$.

\paragraph{Step 2: Purification cost estimation}
Using the per-hop targets from Step 1 and the realized initial fidelities from Phase 3, each node computes the minimum number of symmetric BBPSSW \cite{Bennett96Purification} purification rounds needed on each hop. For initial fidelity $F^{(0)}$ and target $F'$, we iteratively apply
\begin{equation}
  F^{(r+1)} = \frac{F^{(r)^2} + \tfrac{1}{9}(1 - F^{(r)})^2}
               {F^{(r)^2} + \tfrac{2}{3}F^{(r)}(1 - F^{(r)}) + \tfrac{5}{9}(1 - F^{(r)})^2}
  \label{eq:bbpssw}
\end{equation}
to calculate the number of rounds $C_{\mathrm{pur}}(e, F') = \min\{r : F^{(r)} \geq F'\}$. We mark the target as \emph{infeasible} (and stop iteration) if the number of rounds exceeds $R_{\max}$. Each round consumes one additional raw pair, so $2^{C_{\mathrm{pur}}}$ raw pairs are needed to produce one purified output pair on hop $e$. These cost estimates are computed for every hop on the major path and on each candidate recovery path within the node's $k$-hop view.

\paragraph{Step 3: Recovery span selection via EXG}
Q-GUARD triggers recovery when every channel across a reserved link fails to generate entanglement. When recovery is triggered, the algorithm must select among candidate recovery detours whose switch nodes are on either side of the failed segment. To make this selection, we introduce the \emph{expected goodput} (EXG) metric, a segment-local scalar that jointly accounts for swap success, purification overhead, and resource availability.

For a candidate span $P$ with width $W$, $S$ intermediate swaps, and per-hop purification round counts $\{r_e\}_{e \in P}$ from Step 2, EXG is defined as
\begin{equation}
  \mathrm{EXG}(P) = \frac{W \cdot q^{S}}{1 + \sum_{e \in P} (2^{r_e} - 1)} \cdot A_{\min}(P)
  \label{eq:exg}
\end{equation}
where $q^S$ is the probability that all swaps succeed, the denominator penalizes purification overhead (each hop consuming $2^{r_e}$ raw pairs per purified output), and $A_{\min}(P) \in [0,1]$ is the \emph{bottleneck availability factor}: the minimum, across all hops, of the ratio of actually available entangled pairs to the number required by the purification plan. A span is declared infeasible if any hop's target is unreachable within $R_{\max}$ rounds or requires more raw pairs than the path width allows.

A recovery detour replaces a major-path segment only if (i) it is EXG-feasible and (ii) its EXG exceeds that of all other candidate detours for that segment. EXG serves as a local ranking heuristic during recovery planning; final qualification of ebits is determined end-to-end in Phase 5 based on the realized fidelity of the assembled path.

After these steps, each major path has a final route composed of surviving major-path segments and selected recovery detours, along with a purification plan specifying the target fidelity and estimated round count for each hop.

\subsubsection{Phase~5: Purification, Swapping, and Qualification}
\label{sec:phase5}

Nodes execute the purification plan produced by Phase 4 along the assembled route. On each hop, the algorithm repeatedly selects the highest-fidelity available Bell pair for BBPSSW purification by consuming one pair to improve another, until the per-hop target is met or fewer than two pairs remain. Intermediate nodes then perform entanglement swaps using the highest-fidelity available pair on each hop.

After swapping, the source and destination may hold end-to-end Bell pairs that fall slightly below $F_{\mathrm{th}}$, due to minor variations in realized fidelities across hops. To recover these near-threshold pairs, a final end-to-end purification step is applied. Among all assembled end-to-end pairs, those that do not yet meet $F_{\mathrm{th}}$ are iteratively purified by combining the two lowest-fidelity unqualified pairs via BBPSSW, consuming one pair to improve the other. This continues until all remaining pairs meet $F_{\mathrm{th}}$, or fewer than two unqualified pairs remain. All Q-GUARD variants employ this final end-to-end purification step. The resulting end-to-end fidelity is computed using the Werner model ($w_{\mathrm{end}} = \prod_i w(F_i)$), and only Bell pairs with $F_{\mathrm{end}} \geq F_{\mathrm{th}}$ are counted as qualified ebits.

\subsection{Illustrative Example}
\label{sec:example}
\begin{figure}[t]
  \centering
  \begin{minipage}[b]{\columnwidth}
    \centering
    \includegraphics[width=0.85\columnwidth]{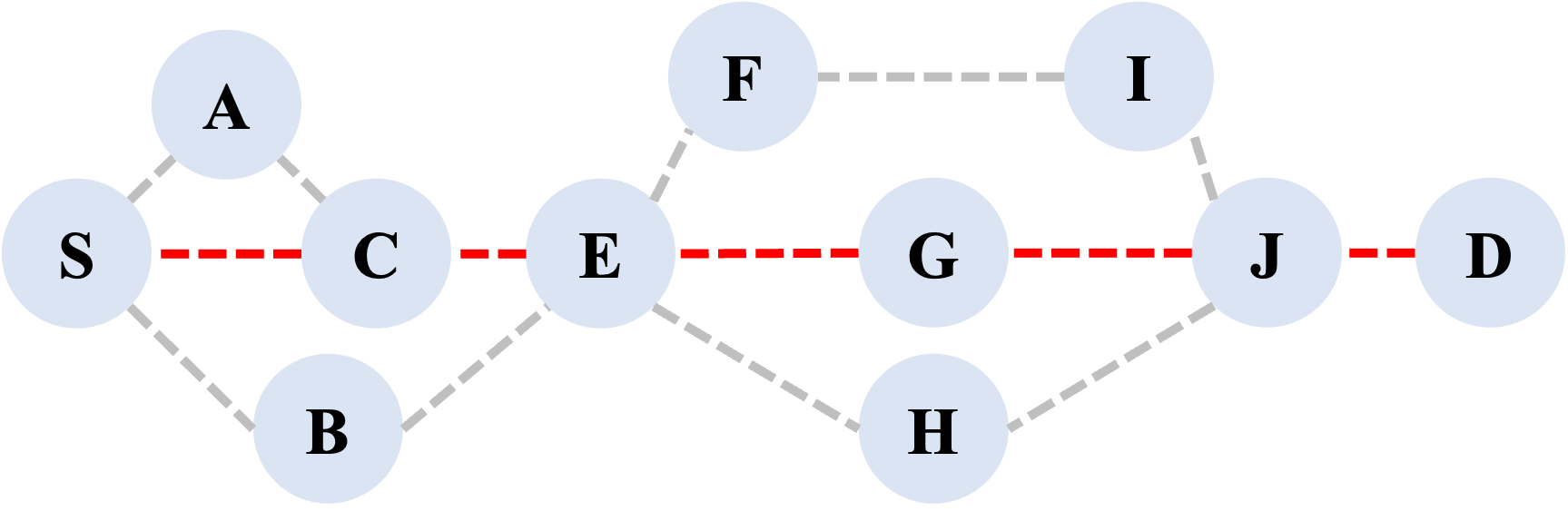}
    \subcaption{Phase~2 selects major path $S$--$C$--$E$--$G$--$J$--$D$ (red dashed) and recovery paths (gray dashed).}
    \label{fig:example-a}
  \end{minipage}\\[6pt]
  \begin{minipage}[b]{\columnwidth}
    \centering
    \includegraphics[width=0.85\columnwidth]{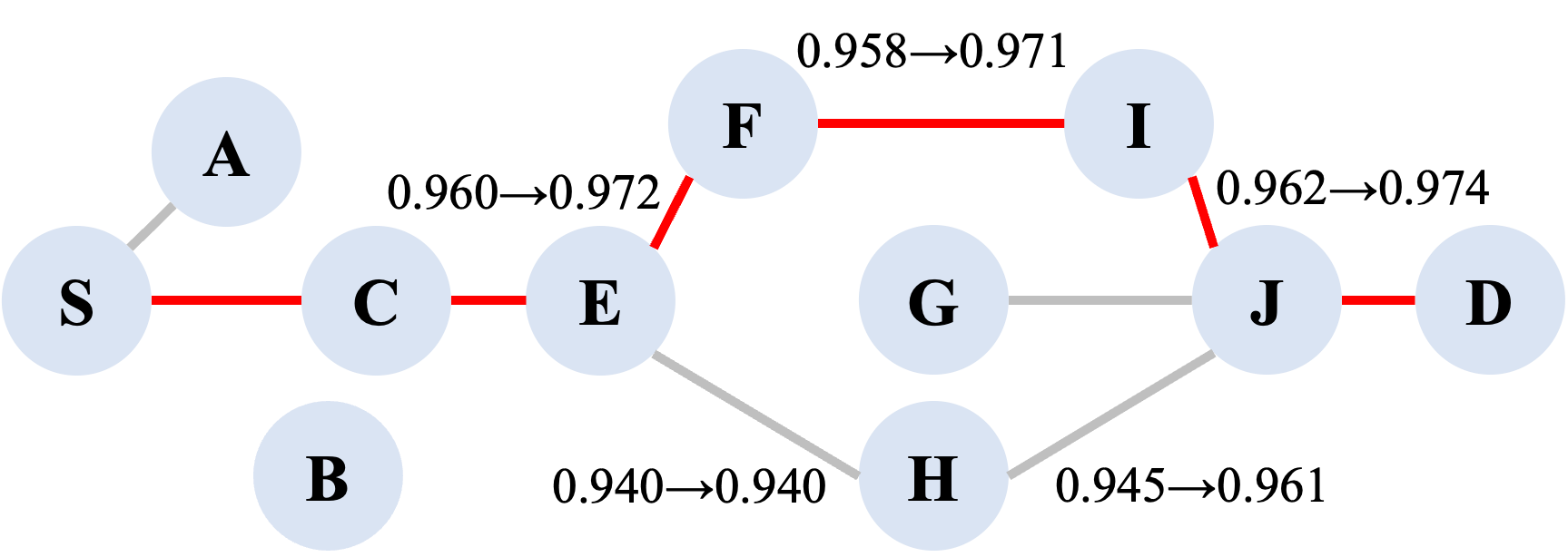}
    \subcaption{After Phase~3, link $E$--$G$ fails. Labels show initial and purified fidelity on relevant links. Q-GUARD selects the 3-hop detour $E$--$F$--$I$--$J$ (red solid), which is fidelity-feasible despite being longer.}
    \label{fig:example-b}
  \end{minipage}
  \caption{Illustrative example of Q-GUARD recovery ($F_{\mathrm{th}} = 0.8$, $L = 5$ hops).}
  \label{fig:example}
\end{figure}

Figure~\ref{fig:example} illustrates Q-GUARD's recovery procedure. In Phase 2, the major path $S$--$C$--$E$--$G$--$J$--$D$ (red, dashed) is selected with $L = 5$ hops, along with 2-hop recovery paths through nodes $A$, $B$, and $H$, and a 3-hop recovery path through segment $F$--$I$ (gray, dashed). In Phase 3, link generation succeeds on all reserved links except $E$--$G$, which fails.

Because $E$--$G$ has failed, Q-GUARD must select a recovery detour. The failed hop lies within the segment $E$--$G$--$J$ ($l = 2$ hops on the major path), and there are two candidate detours: the 2-hop path $E$--$H$--$J$ and the 3-hop path $E$--$F$--$I$--$J$. A throughput-oriented algorithm such as Q-CAST would prefer the shorter 2-hop detour, since its recovery procedure favors fewer intermediate swaps. Q-GUARD, however, evaluates fidelity feasibility before making this decision.

\paragraph{Deriving per-hop targets}
The requested threshold is $F_{\mathrm{th}} = 0.8$, giving $w_{\mathrm{th}} = (4 \times 0.8 - 1)/3 \approx 0.733$ (Eq. \eqref{eq:e2e}). The replaced segment spans $r = 2$ of the major path's $L = 5$ hops, so its Werner budget under the equal-split rule is $w_{\mathrm{seg}} = w_{\mathrm{th}}^{l/L} = 0.733^{2/5} \approx 0.883$ (Eq. \eqref{eq:uniform-target}).

For the \textbf{2-hop detour} ($l' = 2$), each hop's target Werner parameter is $w_{\mathrm{seg}}^{1/l'} = 0.883^{1/2} \approx 0.940$, corresponding to $F_{\mathrm{hop}} \geq 0.955$. For the \textbf{3-hop detour} ($l' = 3$), each hop requires $w_{\mathrm{seg}}^{1/3} \approx 0.960$, corresponding to $F_{\mathrm{hop}} \geq 0.970$---a \emph{higher} per-hop target, because three hops must each contribute a Werner parameter closer to 1 to achieve the same product.

\paragraph{Feasibility evaluation}
Table~\ref{tab:example} summarizes the feasibility of each detour. Post-purification fidelities are computed using the BBPSSW recurrence (Eq.~\eqref{eq:bbpssw}) with symmetric inputs. The 2-hop path through $H$ requires each hop to reach $F \geq 0.955$, but the $E$--$H$ link produced only one successful pair with initial fidelity $F_e^{(0)} = 0.940$. With a single pair, no purification is possible, and the link falls short of its target. The 2-hop detour is therefore infeasible.

The 3-hop path through $F$ and $I$ demands a higher per-hop target of $F \geq 0.970$, but each of its links has enough available pairs to meet this target with one round of BBPSSW purification (Table~\ref{tab:example}). Q-GUARD selects the 3-hop detour, purifies each hop to its target, and performs entanglement swaps at $F$ and $I$, contributing a segment whose Werner product meets the allocated budget $w_{\mathrm{seg}}$.

This example highlights a key property of fidelity-aware routing: \emph{shorter paths are not always better}. A shorter detour requires a lower per-hop target, but it is constrained to use whichever links are available along that route. If any link cannot be purified to its target (as with $E$--$H$ here) the entire detour is infeasible. A longer detour with higher-quality links can succeed despite requiring more from each hop. Q-GUARD's EXG metric captures this tradeoff, while Q-CAST's EXT metric would select the shorter, fidelity infeasible path.

\begin{table}[t]
\small
\centering
\caption{Fidelity feasibility of recovery detours ($F_{\mathrm{th}} = 0.8$, segment budget $w_{\mathrm{seg}} = 0.883$).}
\label{tab:example}
\begin{tabular}{@{}llccccc@{}}
\toprule
Path & Hop & $F_e^{(0)}$ & Pairs & $F_{\mathrm{PUR}}$ & Target $F$ & Feasible? \\
\midrule
\multirow{2}{*}{\shortstack[l]{2-hop}}
  & $E$--$H$ & 0.940 & 1 & 0.940 & 0.955 & \texttimes \\
  & $H$--$J$ & 0.945 & 2 & 0.961 & 0.955 & \checkmark \\
\midrule
\multirow{3}{*}{\shortstack[l]{3-hop}}
  & $E$--$F$ & 0.960 & 2 & 0.972 & 0.970 & \checkmark \\
  & $F$--$I$ & 0.958 & 3 & 0.971 & 0.970 & \checkmark \\
  & $I$--$J$ & 0.962 & 2 & 0.974 & 0.970 & \checkmark \\
\bottomrule
\end{tabular}
\end{table}

\section{Q-GUARD-WS: Weighted Split}
\label{sec:ws}

Q-GUARD's equal-split target allocation (Eq.~\ref{eq:uniform-target}) assigns every hop the same fidelity target regardless of hardware quality. This is robust when per-link quality is unknown, but wastes purification resources when link quality varies: strong links are over-purified while weak links may still fail to meet their target. Q-GUARD-WS addresses this by modifying Phase 4's target allocation to account for hardware heterogeneity. All other phases are unchanged; in particular, Phase 2 still uses EXT-based path scoring identical to Q-CAST and Q-GUARD.

Q-GUARD-WS requires that each link's hardware quality parameter $\eta_e$ is known as part of the static network topology, enabling reliable estimation of initial fidelity via Eq. \eqref{eq:f0}. This is a stronger assumption than base Q-GUARD, which requires no prediction of link quality. In practice, $\eta_e$ could be obtained through periodic calibration or manufacturer specifications; however, as noted in~\cite{Abane25}, per-link fidelity estimation remains non-trivial. Q-GUARD-WS is therefore best suited to deployments where link characterization data is available, while base Q-GUARD provides a more conservative design that operates without this information. It modifies two components of Phase 4: the per-hop target allocation for major paths (Step 1) and the Werner budget assigned to recovery detours.

\subsection{Greedy effort distribution}
\label{sec:greedy-effort}

Given a candidate path of $L$ hops and width $W$, Q-GUARD-WS uses each link's hardware quality parameter $\eta_e$ to estimate its initial fidelity using Eq. \eqref{eq:f0}, then solves for a non-uniform round allocation $\{r_e\}$ that meets the end-to-end threshold $w_{\mathrm{th}}$ while minimizing total purification cost:

\begin{enumerate}[leftmargin=1.5em, itemsep=2pt, topsep=3pt]
    \item For each hop $e$ and $r = 0, 1, \ldots, R_{\max}$, compute the fidelity after $r$ rounds of ideal symmetric BBPSSW, $F_e^{(r)}$, by iterating Eq.~\eqref{eq:bbpssw} from the estimated initial fidelity.
    \item Find the minimum uniform round count $R$ such that $\prod_e w(F_e^{(R)}) \geq w_{\mathrm{th}}$. If no such $R \leq R_{\max}$ exists or $2^R > W$, the path is infeasible.
    \item Starting from $r_e = R$ for all hops, greedily decrement one round at a time: among all hops where $r_e > 0$ and decrementing still satisfies $\prod_e w(F_e^{(r_e)}) \geq w_{\mathrm{th}}$, choose the hop with the smallest reduction in fidelity $w(F_e^{(r_e)}) - w(F_e^{(r_e - 1)})$ and decrement its round count. Repeat until no further decrements are feasible.
\end{enumerate}

This yields a round allocation that meets the end-to-end fidelity threshold while minimizing total purification cost. Hops with naturally high fidelity perform fewer rounds, as their rounds have less marginal gain, while hops with low initial fidelity---and therefore a larger fidelity gain per round---receive more purification effort.

\subsection{Hardware-weighted recovery budgets}
\label{sec:ws-recovery}

Q-GUARD's equal-split rule assigns each recovery detour a Werner budget proportional to the number of major-path hops it replaces: $w_{\mathrm{seg}} = w_{\mathrm{th}}^{l/L}$ for a segment of $l$ hops on an $L$ hop major path. This treats all hops as equally difficult, regardless of hardware quality.

Q-GUARD-WS instead allocates the budget proportional to the aggregate depolarization burden of the replaced segment:
\begin{equation}
  w_{\mathrm{seg}} = w_{\mathrm{th}}^{\,D_{\mathrm{seg}} / D_{\mathrm{total}}}
  \label{eq:ws-seg}
\end{equation}
where $D_{\mathrm{seg}} = \sum_{e \in \mathrm{seg}} 1/\eta_e$ is the total depolarization weight of the replaced segment and $D_{\mathrm{total}} = \sum_{e \in P} 1/\eta_e$ is the total for the major path. Links with lower $\eta_e$ (worse hardware) contribute more depolarization weight, so segments containing weak links receive a proportionally larger share of the Werner budget. The detour's per-hop round allocation is then computed using the greedy effort distribution (Section~\ref{sec:greedy-effort}) applied to the recovery path with the budget from Eq.~\eqref{eq:ws-seg}.

\section{Q-GUARD-FP: Fidelity-Aware Path Selection}
\label{sec:fp}

As an exploratory variant, we evaluate whether incorporating predicted link quality into Phase 2 path \emph{selection} can improve performance beyond Q-GUARD's Phase 4 purification planning alone. Q-GUARD-FP uses the same hardware quality parameters $\eta_e$ as Q-GUARD-WS to replace Q-CAST's EXT scoring function in Phase 2 with a purification-aware generalization. All other phases, including Phase 4's fidelity-aware recovery planning and Phase 5's purification execution, remain identical to base Q-GUARD.

\subsection{Generalizing EXT to purification-aware scoring}

Q-CAST's EXT metric (Eqs. \ref{eq:ext-q}--\ref{eq:ext}) tracks the bottleneck distribution of \emph{raw} successful links across hops. Q-GUARD-FP generalizes this to track \emph{purified output pairs} by modifying the per-hop distribution $Q_k^i$.

Given the end-to-end threshold $F_{\mathrm{th}}$ and path length $L$, each hop is assigned a target fidelity via the equal-split rule (Eq. \ref{eq:uniform-target}). Using the predicted initial fidelity from $\eta_e$ (Eq. \ref{eq:f0}), the minimum number of BBPSSW rounds $r_k$ is computed from Eq. \eqref{eq:bbpssw}, giving a purification cost of $c_k = 2^{r_k}$ raw pairs per purified output. Define $M = \min_k \lfloor W / c_k \rfloor$ as the maximum purified bottleneck width; if $M = 0$ or any hop's target is unreachable, the path receives a score of zero.

The per-hop distribution $Q_k^i$ from Eq.~\eqref{eq:ext-q} is replaced by the probability that hop~$k$ produces exactly $i$ \emph{purified} pairs:
\begin{equation}
  Q_k^{\prime\,i} =
    \sum_{j \,=\, i \cdot c_k}^{\min\bigl((i+1)\,c_k - 1,\; W\bigr)}
    \binom{W}{j}\, p_k^{\,j}\,(1 - p_k)^{W - j}
  \label{eq:fp-q}
\end{equation}
for $i = 0, 1, \ldots, M{-}1$, with $Q_k^{\prime\,M}$ absorbing the remaining tail. The bottleneck recurrence (Eq.~\ref{eq:ext-bottleneck}) and final score (Eq.~\ref{eq:ext}) then proceed identically, with $Q_k^{\prime\,i}$ replacing $Q_k^i$ and $M$ replacing $W$:
\begin{equation}
  \mathrm{score}(P) = q^{\,L-1} \cdot \sum_{i=1}^{M} i \cdot P_L^{\prime\,i}.
  \label{eq:fp-score}
\end{equation}

When no purification is needed ($r_k = 0$ for all $k$), $c_k = 1$ and $M = W$, so $Q_k^{\prime\,i}$ reduces to $Q_k^i$ and the score becomes EXT. This scoring function thus strictly generalizes EXT to account for the width reduction imposed by purification. 

The Extended Dijkstra Algorithm requires that the scoring function decreases monotonically as a path extends \cite{QCAST}. This is preserved because extending a path can only reduce $M$ (due to increased per-hop costs as $L$ grows) and introduces an additional swap factor $q$.

\subsection{Discussion}

Q-GUARD-FP is presented as an exploratory variant rather than a recommended configuration. As we show in Section~\ref{sec:fp-results}, incorporating predicted link quality into Phase 2 scoring provides no meaningful throughput advantage over base Q-GUARD while degrading effective service range, validating Q-GUARD's choice of deferring fidelity decisions to Phase 4.

\section{Results}
\label{sec:results}

We evaluate Q-GUARD and Q-GUARD-WS against two baselines:
(i) \textbf{Q-CAST}~\cite{QCAST}, the original distributed throughput-oriented algorithm, and
(ii) \textbf{Q-CAST-PUR}, a strengthened variant that runs Q-CAST unchanged and then applies end-to-end BBPSSW purification to the delivered Bell pairs by repeatedly combining the two lowest-fidelity unqualified pairs until all remaining pairs meet $F_{\mathrm{th}}$ or fewer than two pairs remain. Q-CAST-PUR represents the most direct approach to add fidelity enforcement to Q-CAST without modifying path selection or recovery decisions.

\begin{figure*}[!b]
  \centering
  \begin{minipage}[b]{0.24\textwidth}
    \centering
    \includegraphics[width=\textwidth]{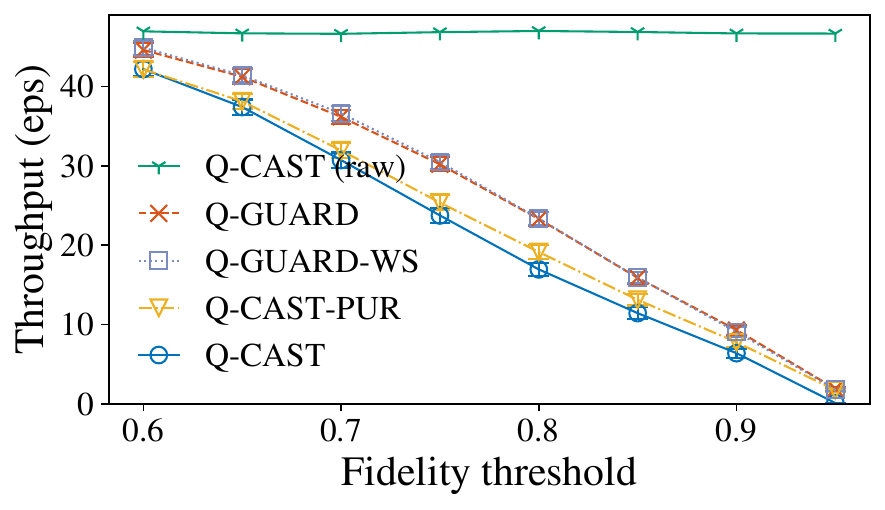}
    \subcaption{}
    \label{fig:throughput-fth}
  \end{minipage}\hfill
  \begin{minipage}[b]{0.24\textwidth}
    \centering
    \includegraphics[width=\textwidth]{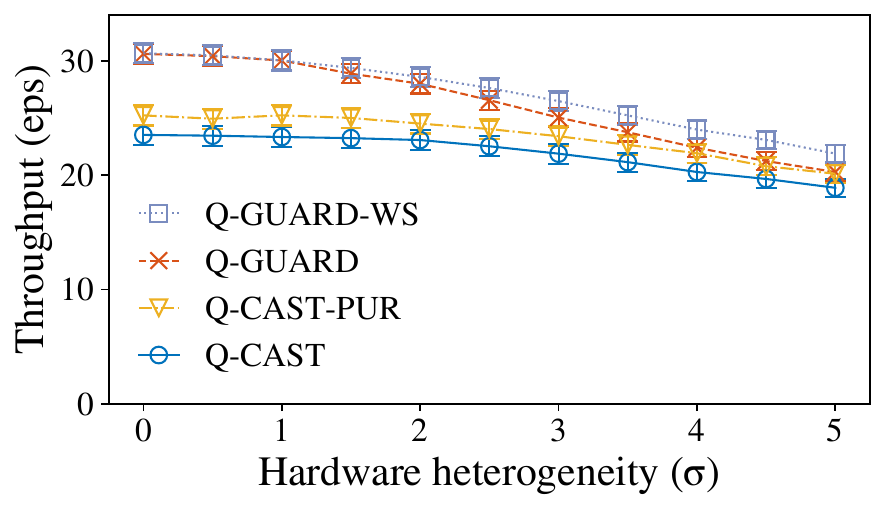}
    \subcaption{}
    \label{fig:throughput-eta}
  \end{minipage}\hfill
  \begin{minipage}[b]{0.24\textwidth}
    \centering
    \includegraphics[width=\textwidth]{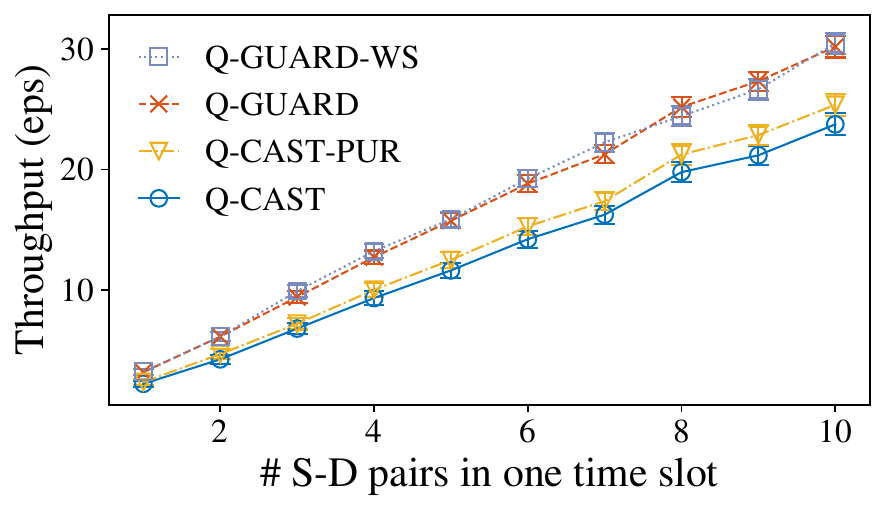}
    \subcaption{}
    \label{fig:throughput-nsd}
  \end{minipage}\hfill
  \begin{minipage}[b]{0.24\textwidth}
    \centering
    \includegraphics[width=\textwidth]{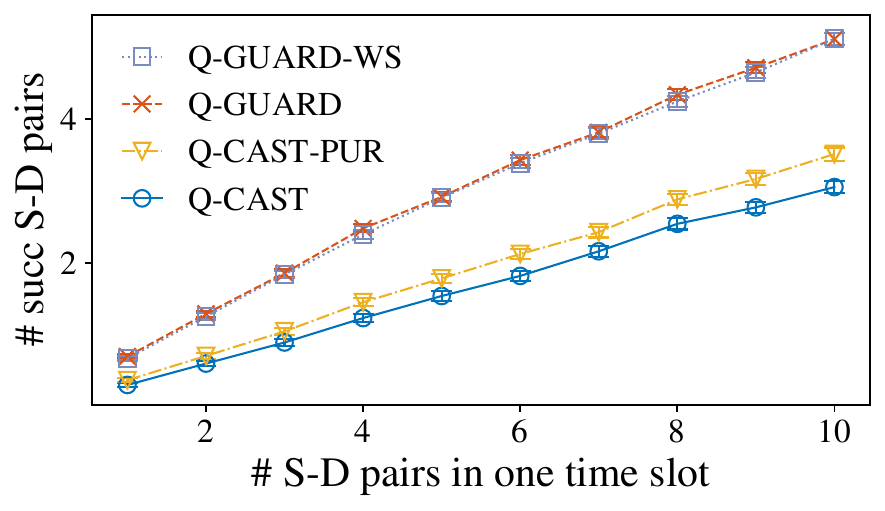}
    \subcaption{}
    \label{fig:pairs-nsd}
  \end{minipage}
  \caption{Performance across operating conditions
  (a)~Qualified throughput vs.\ fidelity threshold. Q-CAST's raw throughput (green) is independent of $F_{\mathrm{th}}$.
  (b)~Qualified throughput vs.\ hardware heterogeneity $\sigma_\eta$.
  (c)~Qualified throughput vs.\ number of offered S--D pairs.
  (d)~Qualified S--D pairs served vs.\ number of offered S--D pairs.}
  \label{fig:sweep}
\end{figure*}

\subsection{Experimental setup}
\label{sec:setup}
We generate random connected topologies using a Waxman model \cite{Waxman88, Pant20} on a 100\,km $\times$ 100\,km area, following the same approach as Q-CAST \cite{QCAST} which produces realistic metropolitan-area topologies with heterogeneous link lengths. Unless otherwise noted, the reference settings for experiments use $|V| = 100$ nodes, average degree $d = 6$, average channel success probability $p = 0.7$, swap success probability $q = 0.9$, link-state broadcast radius $k = 3$ hops, $m = 10$ concurrent S--D pairs per slot, and fidelity threshold $F_{\mathrm{th}} = 0.75$. Each node has 20--31 qubit memories, and each edge supports 6--12 parallel channels, both drawn uniformly at random. Hardware quality parameters are drawn independently as $\eta_e \sim \mathcal{N}(9.5,\, 1.0)$, producing heterogeneous initial fidelities centered around $F_e^{(0)} \approx 0.925$ with most links falling in the range 0.91--0.94 (Eq.~\eqref{eq:f0}), plus per-generation Gaussian noise ($\sigma = 0.01$). For each parameter configuration, we simulate 1000 independent time slots and report metric averages.

We report two primary metrics:
\begin{itemize}[leftmargin=1.5em,itemsep=1pt,topsep=2pt]
  \item \textbf{Qualified throughput} (ebits per slot, eps): the total number of end-to-end Bell pairs per time slot whose end-to-end fidelity meets $F_{\mathrm{th}}$.
  \item \textbf{Qualified S--D pairs}: the number of distinct request pairs that receive at least one qualified ebit in a slot.
\end{itemize}
The first metric captures total service volume; the second captures how the service is distributed across concurrent flows.
%---a measure more relevant to applications that require at least one qualified ebit per session.

\subsection{Qualified Throughput and Service Coverage}
\label{sec:sweep}

Figure~\ref{fig:sweep} evaluates Q-GUARD across four operating dimensions. All experiments use the reference setting unless the varied parameter is noted.

\paragraph{Throughput vs. fidelity threshold}
Figure \ref{fig:sweep}(a) varies the requested fidelity threshold $F_{\mathrm{th}}$ from 0.60 to 0.95. For reference, Q-CAST's \emph{raw} throughput (all delivered pairs, regardless of fidelity) is shown at approximately 48 eps. This is the ceiling that our fidelity-aware algorithm is bound by, since purification consumes raw pairs and fidelity enforcement rejects sub-threshold ones.

At low thresholds ($F_{\mathrm{th}} \leq 0.65$), all algorithms convert most delivered pairs into qualified ebits, with Q-GUARD and Q-GUARD-WS achieving roughly 43--46 of Q-CAST's 48 raw eps. As $F_{\mathrm{th}}$ increases, Q-CAST's qualified throughput drops steeply as its delivered pairs increasingly fall below threshold. Q-CAST-PUR degrades more gracefully, since its end-to-end purification salvages some near-threshold pairs. Q-GUARD and Q-GUARD-WS maintain a consistent advantage through the per-hop purification planning: at $F_{\mathrm{th}} = 0.75$, they deliver approximately 31 qualified pairs compared to roughly 26 for Q-CAST-PUR and 24 for Q-CAST (an increase of about 30\%). At very high thresholds ($F_{\mathrm{th}} \geq 0.90$), all algorithms converge as the physical limits of the link model are approached.

\paragraph{Throughput vs. hardware heterogeneity}
Figure \ref{fig:sweep}(b) varies the hardware heterogeneity ($\sigma_\eta$) from 0 (all links identical) to 5 (high variability in link quality). At low heterogeneity, Q-GUARD and Q-GUARD-WS perform identically, as equal-split and weighted-split allocations converge when all links have similar quality. As $\sigma_\eta$ increases, Q-GUARD-WS starts to outperform Q-GUARD: at $\sigma_\eta = 5$, Q-GUARD-WS delivers approximately 22 qualified eps compared to 20 for Q-GUARD. This confirms the design rationale of Q-GUARD-WS: by concentrating effort on weak links and reducing unnecessary effort on strong ones, the greedy effort distribution reduces wasted purification resources. All four algorithms decline with increasing heterogeneity, as more links fall below the quality needed for reliable service. 

\paragraph{Throughput vs. offered load}
Figure \ref{fig:sweep}(c) varies the number of requested concurrent S--D pairs from 1 to 10. All algorithms scale roughly linearly, but the Q-GUARD variants maintain a consistent advantage. At $m = 10$, Q-GUARD produces approximately 30 qualified eps compared to roughly 25 for Q-CAST-PUR and 23 for Q-CAST. The gap widens as load increases, confirming that Q-GUARD's fidelity aware resource allocation becomes more valuable under contention.

\paragraph{Service coverage vs. offered load}
Figure~\ref{fig:sweep}(d) shifts from aggregate throughput to service coverage: the number of distinct S--D pairs receiving at least one qualified ebit. 
% This metric is relevant to applications such as QKD sessions that require at least one qualified ebit to proceed. 
At $m = 10$, Q-GUARD and Q-GUARD-WS serve approximately 5.1 S--D pairs per slot, compared to roughly 3.5 for Q-CAST-PUR and 3.0 for Q-CAST---a 70\% increase over Q-CAST. This shows that under load, several requests receive no qualified ebits at all.
This motivates the range analysis in Section \ref{sec:range}: the unserved pairs are predominantly those whose S--D distance exceeds the effective service radius.

\subsection{Range Extension}
\label{sec:range}

\begin{figure}[t]
  \centering
  \includegraphics[width=0.75\columnwidth]{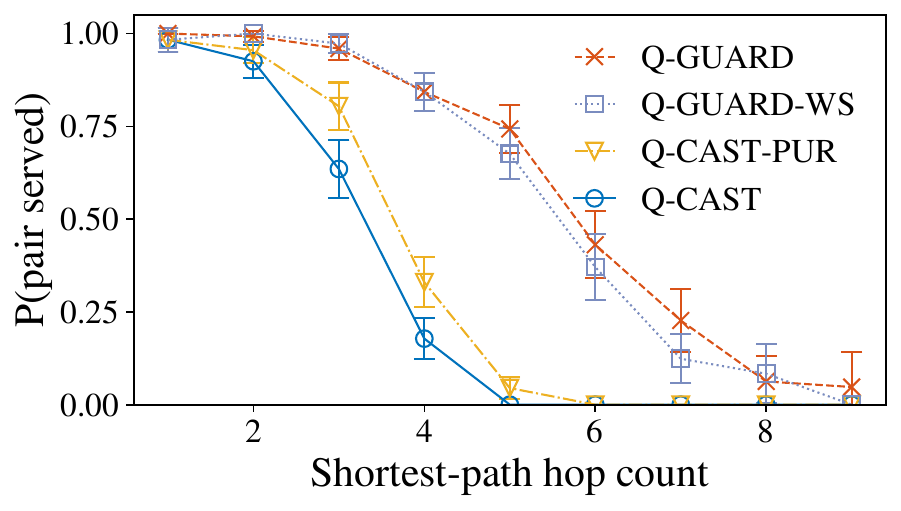}
  \caption{Probability that an S--D pair receives at least one qualified ebit vs.\ shortest-path hop count ($|V|{=}100$, $m{=}1$,
  $F_{\mathrm{th}}{=}0.75$).}
  \label{fig:range-hops}
\end{figure}

\begin{figure}[t]
  \centering
  \includegraphics[width=0.75\columnwidth]{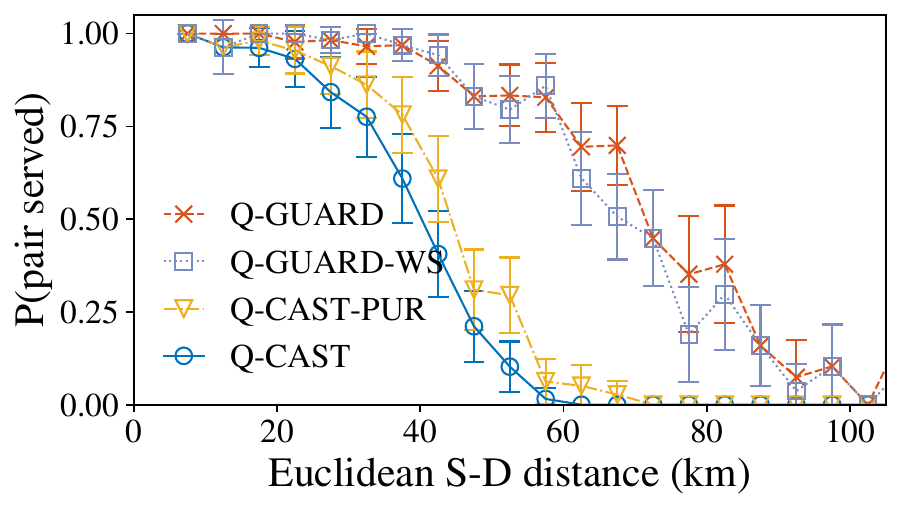}
  \caption{Probability that an S--D pair receives at least one qualified ebit vs.\ Euclidean S--D distance ($|V|{=}100$, $m{=}1$,
  $F_{\mathrm{th}}{=}0.75$).}
  \label{fig:range-dist}
\end{figure}

The most significant result of this evaluation is Q-GUARD's extension of the effective range over which fidelity-qualified ebits can be reliably delivered. To isolate distance-dependent behavior from contention, we run separate experiments using the same reference settings but with a single, random S--D pair per slot. Figures~\ref{fig:range-hops} and~\ref{fig:range-dist} plot the fraction of time slots in which the pair receives at least one qualified ebit, binned by shortest-path hop count and Euclidean S--D distance, respectively.

\paragraph{Range by hop count}
In Figure \ref{fig:range-hops}, Q-CAST's qualified success rate drops sharply after 2 hops, falling below 20\% at 4 hops and reaching zero by 5 hops. Q-CAST-PUR extends this modestly, maintaining roughly 35\% at 4 hops, through its end-to-end purification, but also reaching zero by 6 hops. Q-GUARD achieves over 85\% at 4 hops and approximately 75\% at 5 hops, where both baselines have effectively failed. At 6 hops, Q-GUARD still delivers qualified pairs for more than 45\% of requests, with nonzero success rates extending to 9 hops. Q-GUARD-WS is somewhat weaker in the 4--5 hop range, but is generally comparable in most regimes.

\paragraph{Range by Euclidean distance}
Figure \ref{fig:range-dist} offers the same analysis in physical units. Q-CAST falls below the 50\% qualified success threshold at roughly 40 km and Q-CAST-PUR at approximately 45 km, while Q-GUARD reaches it at roughly 75 km, nearly doubling the effective service radius. For applications requiring reliable fidelity-qualified service across a metropolitan-scale network (50--100 km), Q-GUARD substantially extends the feasible operating range.

\subsection{Q-GUARD-FP Evaluation}
\label{sec:fp-results}

To isolate the effect of purification-aware Phase 2 path scoring (Section~\ref{sec:fp}), we compare Q-GUARD-FP against base Q-GUARD. Across most metrics, including qualified throughput as a function of fidelity threshold, hardware heterogeneity, and offered load Q-GUARD-FP performs comparably to Q-GUARD, with at most a marginal improvement of 0.5--1 eps at low to moderate heterogeneity levels.

\begin{figure}[t]
  \centering
  \includegraphics[width=0.75\columnwidth]{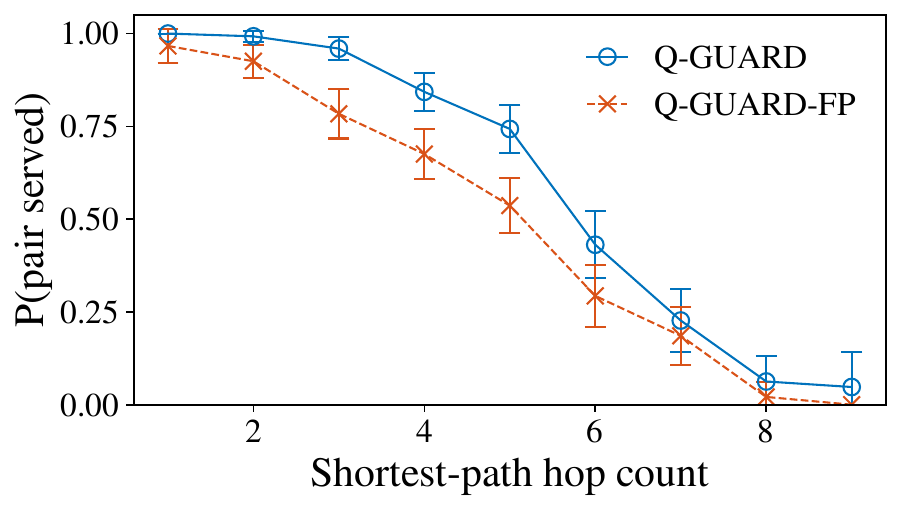}
  \caption{Q-GUARD-FP vs.\ Q-GUARD: qualified success rate vs.\
  shortest-path hop count ($|V|{=}100$, $m{=}1$,
  $F_{\mathrm{th}}{=}0.75$).}
  \label{fig:fp-range}
\end{figure}

However, Figure~\ref{fig:fp-range} reveals a substantial cost in service range. Q-GUARD-FP's qualified success rate is lower than Q-GUARD's at nearly all hop counts, with the gap widest around 4--5 hops---where Q-GUARD achieves roughly 85\% while Q-GUARD-FP reaches only about 67\%. This is because Q-GUARD-FP assigns a score of zero to paths whose \emph{predicted} purification cost exceeds the available width, rejecting paths that may succeed in practice when realized fidelities (which include per-generation noise) turn out better than predicted, or when Phase 4's adaptive planning compensates for individual hops that fall slightly short. Base Q-GUARD defers all fidelity decisions to Phase 4, giving it a larger candidate path set.

This tradeoff illustrates why we present Q-GUARD-FP only as an exploration rather than a recommended variant: the throughput gain is marginal, while the range degradation is substantial. For metropolitan-scale networks where extending the service range is a primary objective, Q-GUARD's EXT-based Phase 2 scoring provides a better balance.

\section{Conclusion}
\label{sec:conclusion}

We presented Q-GUARD, a distributed entanglement routing algorithm that provides per-request, end-to-end fidelity guarantees within a distributed, $k$-hop local link-state model. Q-GUARD preserves Q-CAST's contention-aware path selection and recovery structure, but adds a fidelity-aware planning phase that allocates per-hop targets from a requested threshold $F_{\text{th}}$, constructs per-link purification cost tables from the raw Bell pairs generated in each slot, and uses a segment-local expected goodput (EXG) metric to choose between a major-path segment and its candidate recovery detours. Q-GUARD-WS extends this with a greedy effort distribution that exploits per-link initial fidelity estimates to allocate purification rounds non-uniformly, concentrating effort on hops that need it and skipping purification on hops with naturally high fidelity. As an exploratory variant, we also derived a purification-aware generalization of Q-CAST's optimal EXT path metric for Phase 2 scoring (Q-GUARD-FP), finding that using predicted link quality in path selection yields marginal throughput gains but degrades effective range. This validates Q-GUARD's choice of deferring fidelity decisions to Phase 4. 

Our evaluation on synthetic 100-node topologies demonstrates three key findings. First, Q-GUARD consistently delivers higher fidelity-qualified throughput and serves more S--D pairs than both baselines across a range of operating conditions, producing approximately 30\% more qualified ebits than Q-CAST at the reference setting. Second, and most significantly, Q-GUARD nearly doubles the effective fidelity-qualified service radius, maintaining over 85\% success at 4 hops where Q-CAST falls below 20\%. Third, Q-GUARD-WS provides additional gains that grow with hardware heterogeneity, making it well-suited to deployments with mixed-generation or multi-vendor equipment.

This work has several limitations that point to future directions. We rely on a single Werner-state noise model and homogeneous swap probabilities; extending Q-GUARD to device-specific noise models and heterogeneous hardware is a natural next step. Our simulation model does not model multi-slot memory management or time-based decoherence, which would affect purification planning in networks with longer coherence times. Finally, we evaluate on synthetic topologies only; validation on realistic network topologies and hardware-calibrated simulators such as NetSquid or SeQUeNCe \cite{Coopmans21NetSquid,Wu21SeQUeNCe} would strengthen confidence in the results. Recent experimental progress---such as metropolitan-scale memory--memory entanglement~\cite{Liu24, stolk2024metropolitan, craddock2026high}---suggests that the distance regimes targeted by Q-GUARD are becoming physically realizable. More broadly, we hope that the ideas in this work (per-slot fidelity qualification, span-level purification planning, and lightweight metrics like EXG) can inform the design of fidelity-aware routing in emerging quantum network testbeds.

\appendix
\section{Computational Overhead}
\label{sec:overhead}

Q-GUARD's fidelity-aware planning adds computation beyond Q-CAST's. We analyze the additional cost using the notation of \cite{QCAST}: $h_m$ is the max path hop count, $W_m$ is the max path width, $n = |V|$, and $m$ is the number of S--D pairs.

\paragraph{Phase 4: Cost table construction}
Q-GUARD's purification tables are confined to the local $k$-hop link-state view exchanged after link generation. If the maximum node degree is $d$ and the communication distance is $k$, then a node needs to consider at most $O(d \cdot k)$ relevant links when constructing purification cost tables. For each such link, computing the minimum purification rounds requires iterating the BBPSSW recurrence (Eq.~\eqref{eq:bbpssw}) up to $R_{\max}$ times. The resulting complexity is therefore
$O(d \cdot k \cdot R_{\max})$. In practice, $d$, $k$, and $R_{\max}$ are all small constants, so this overhead is modest.

\paragraph{Phase 4: EXG evaluation}
Once the purification cost tables have been constructed, evaluating EXG (Eq.~\eqref{eq:exg}) requires scanning at most the same $O(d \cdot k)$ locally visible links to account for purification costs and availability constraints on candidate spans. Thus the complexity of EXG evaluation is $O(d \cdot k)$, which is asymptotically lower than the cost of cost-table construction.

\paragraph{Phase 4: Q-GUARD-WS greedy effort distribution} 
Q-GUARD-WS adds the greedy effort distribution (Section~\ref{sec:greedy-effort}) to Phase 4's target allocation. For each candidate path, this requires (i) precomputing $F_e^{(r)}$ for each hop and $r = 0, \ldots, R_{\max}$, costing $O(L \cdot R_{\max})$; (ii) finding the minimum feasible uniform round count, costing $O(L \cdot R_{\max})$; and (iii)~the greedy decrement loop, which performs at most $L \cdot R_{\max}$ decrements each requiring an $O(L)$ scan, for a worst-case cost of $O(L^2 \cdot R_{\max})$. Since $L \leq h_m$ and $R_{\max}$ is a small constant, this adds $O(h_m^2)$ per candidate path in Phase 4.

\paragraph{Phase 2: Q-GUARD-FP scoring (exploratory variant only)}
Q-GUARD-FP replaces EXT with the purification-aware score (Eq. \eqref{eq:fp-score}) in EDA. Each path expansion requires recomputing the per-hop purification rounds $\{r_k\}$, costing $O(L \cdot R_{\max})$, and evaluating the full bottleneck recurrence (Eq. \eqref{eq:ext-bottleneck}) over $L$~hops and $M \leq W_m$ purified-width bins, costing $O(L \cdot W_m)$ per expansion. Unlike EXT, the equal-split target changes with path length, so the recurrence cannot be updated incrementally and must be recomputed from scratch at each expansion. The total cost is $O(n \log n + |E| \cdot h_m \cdot (W_m + R_{\max}))$. Since $R_{\max}$ is small, this preserves the asymptotic structure of Q-CAST's EDA.

\paragraph{Summary}
The additional Phase 4 overhead introduced by Q-GUARD is local to the $k$-hop neighborhood and does not require any new global path search after link generation. Its dominant post-generation cost is purification cost-table construction at $O(d\cdot k \cdot R_{\max})$, followed by EXG evaluation at $O(d \cdot k)$. Q-GUARD-WS adds up to $O(L^2 \cdot R_{\max})$ per candidate span for weighted target allocation, while Q-GUARD-FP increases Phase 2 scoring cost but preserves the overall EDA scaling structure. Since $d$, $k$, and $R_{\max}$ are small constants in practice, these additions remain lightweight. 

\bibliographystyle{IEEEtran}
\bibliography{refs}

\begin{thebibliography}{10}
\providecommand{\url}[1]{#1}
\csname url@samestyle\endcsname
\providecommand{\newblock}{\relax}
\providecommand{\bibinfo}[2]{#2}
\providecommand{\BIBentrySTDinterwordspacing}{\spaceskip=0pt\relax}
\providecommand{\BIBentryALTinterwordstretchfactor}{4}
\providecommand{\BIBentryALTinterwordspacing}{\spaceskip=\fontdimen2\font plus
\BIBentryALTinterwordstretchfactor\fontdimen3\font minus \fontdimen4\font\relax}
\providecommand{\BIBforeignlanguage}[2]{{%
\expandafter\ifx\csname l@#1\endcsname\relax
\typeout{** WARNING: IEEEtran.bst: No hyphenation pattern has been}%
\typeout{** loaded for the language `#1'. Using the pattern for}%
\typeout{** the default language instead.}%
\else
\language=\csname l@#1\endcsname
\fi
#2}}
\providecommand{\BIBdecl}{\relax}
\BIBdecl

\bibitem{Kimble08}
H.~J. Kimble, ``The quantum internet,'' \emph{Nature}, vol. 453, no. 7198, p. 1023–1030, Jun 2008.

\bibitem{WehnerRoadmap}
\BIBentryALTinterwordspacing
S.~Wehner, D.~Elkouss, and R.~Hanson, ``Quantum internet: A vision for the road ahead,'' \emph{Science}, vol. 362, no. 6412, p. eaam9288, 2018. [Online]. Available: \url{https://www.science.org/doi/abs/10.1126/science.aam9288}
\BIBentrySTDinterwordspacing

\bibitem{RFC9340}
\BIBentryALTinterwordspacing
W.~Kozlowski, S.~Wehner, R.~V. Meter, B.~Rijsman, A.~S. Cacciapuoti, M.~Caleffi, and S.~Nagayama, ``{Architectural Principles for a Quantum Internet},'' RFC 9340, Mar. 2023. [Online]. Available: \url{https://www.rfc-editor.org/info/rfc9340}
\BIBentrySTDinterwordspacing

\bibitem{Zukowski93Swapping}
M.~Zukowski, A.~Zeilinger, M.~Horne, and A.~Ekert, ``‘‘event-ready-detectors’’ bell experiment via entanglement swapping,'' \emph{Physical review letters}, vol.~71, pp. 4287--4290, 01 1994.

\bibitem{Briegel98}
\BIBentryALTinterwordspacing
H.-J. Briegel, W.~D\"ur, J.~I. Cirac, and P.~Zoller, ``Quantum repeaters: The role of imperfect local operations in quantum communication,'' \emph{Phys. Rev. Lett.}, vol.~81, pp. 5932--5935, Dec 1998. [Online]. Available: \url{https://link.aps.org/doi/10.1103/PhysRevLett.81.5932}
\BIBentrySTDinterwordspacing

\bibitem{Ekert91}
\BIBentryALTinterwordspacing
A.~K. Ekert, ``Quantum cryptography based on bell's theorem,'' \emph{Phys. Rev. Lett.}, vol.~67, pp. 661--663, Aug 1991. [Online]. Available: \url{https://link.aps.org/doi/10.1103/PhysRevLett.67.661}
\BIBentrySTDinterwordspacing

\bibitem{RFC9583}
\BIBentryALTinterwordspacing
C.~Wang, A.~Rahman, R.~Li, M.~Aelmans, and K.~Chakraborty, ``{Application Scenarios for the Quantum Internet},'' RFC 9583, Jun. 2024. [Online]. Available: \url{https://www.rfc-editor.org/info/rfc9583}
\BIBentrySTDinterwordspacing

\bibitem{EFiRAP}
Y.~Zhao, G.~Zhao, and C.~Qiao, ``E2e fidelity aware routing and purification for throughput maximization in quantum networks,'' in \emph{IEEE INFOCOM 2022 - IEEE Conference on Computer Communications}, 2022, pp. 480--489.

\bibitem{QLEAP}
\BIBentryALTinterwordspacing
J.~Li, M.~jian Wang, K.~Xue, R.~Li, N.~Yu, Q.~Sun, and J.~Lu, ``Fidelity-guaranteed entanglement routing in quantum networks,'' \emph{IEEE Transactions on Communications}, vol.~70, pp. 6748--6763, 2021. [Online]. Available: \url{https://api.semanticscholar.org/CorpusID:245335421}
\BIBentrySTDinterwordspacing

\bibitem{QCAST}
\BIBentryALTinterwordspacing
S.~Shi and C.~Qian, ``Concurrent entanglement routing for quantum networks: Model and designs,'' in \emph{Proceedings of the Annual Conference of the ACM Special Interest Group on Data Communication on the Applications, Technologies, Architectures, and Protocols for Computer Communication}, ser. SIGCOMM '20.\hskip 1em plus 0.5em minus 0.4em\relax New York, NY, USA: Association for Computing Machinery, 2020, p. 62–75. [Online]. Available: \url{https://doi.org/10.1145/3387514.3405853}
\BIBentrySTDinterwordspacing

\bibitem{pompili2021realization}
M.~Pompili, S.~L. Hermans, S.~Baier, H.~K. Beukers, P.~C. Humphreys, R.~N. Schouten, R.~F. Vermeulen, M.~J. Tiggelman, L.~dos Santos~Martins, B.~Dirkse \emph{et~al.}, ``Realization of a multinode quantum network of remote solid-state qubits,'' \emph{Science}, vol. 372, no. 6539, pp. 259--264, 2021.

\bibitem{Liu24}
J.-L. {Liu}, X.-Y. {Luo}, Y.~{Yu}, C.-Y. {Wang}, B.~{Wang}, Y.~{Hu}, J.~{Li}, M.-Y. {Zheng}, B.~{Yao}, Z.~{Yan}, D.~{Teng}, J.-W. {Jiang}, X.-B. {Liu}, X.-P. {Xie}, J.~{Zhang}, Q.-H. {Mao}, X.~{Jiang}, Q.~{Zhang}, X.-H. {Bao}, and J.-W. {Pan}, ``{Creation of memory-memory entanglement in a metropolitan quantum network},'' \emph{Nature}, vol. 629, no. 8012, pp. 579--585, May 2024.

\bibitem{stolk2024metropolitan}
A.~J. Stolk, K.~L. van~der Enden, M.-C. Slater, I.~te~Raa-Derckx, P.~Botma, J.~Van~Rantwijk, J.~B. Biemond, R.~A. Hagen, R.~W. Herfst, W.~D. Koek \emph{et~al.}, ``Metropolitan-scale heralded entanglement of solid-state qubits,'' \emph{Science advances}, vol.~10, no.~44, p. eadp6442, 2024.

\bibitem{bersin2024development}
E.~Bersin, M.~Grein, M.~Sutula, R.~Murphy, Y.~Q. Huan, M.~Stevens, A.~Suleymanzade, C.~Lee, R.~Riedinger, D.~J. Starling \emph{et~al.}, ``Development of a boston-area 50-km fiber quantum network testbed,'' \emph{Physical Review Applied}, vol.~21, no.~1, p. 014024, 2024.

\bibitem{craddock2026high}
A.~N. Craddock, T.~Cowan, N.~Bigagli, S.~Yekasiri, D.~Robinson, G.~B. Portmann, Z.~Guo, M.~Kilzer, J.~Zhao, M.~Flament \emph{et~al.}, ``High-rate scalable entanglement swapping between remote entanglement sources on deployed new york city fibers,'' \emph{arXiv preprint arXiv:2602.15653}, 2026.

\bibitem{Bennett96Purification}
\BIBentryALTinterwordspacing
C.~H. Bennett, G.~Brassard, S.~Popescu, B.~Schumacher, J.~A. Smolin, and W.~K. Wootters, ``Purification of noisy entanglement and faithful teleportation via noisy channels,'' \emph{Phys. Rev. Lett.}, vol.~76, pp. 722--725, Jan 1996. [Online]. Available: \url{https://link.aps.org/doi/10.1103/PhysRevLett.76.722}
\BIBentrySTDinterwordspacing

\bibitem{Abane25}
\BIBentryALTinterwordspacing
A.~Abane, M.~Cubeddu, V.~S. Mai, and A.~Battou, ``\BIBforeignlanguage{en}{Entanglement routing in quantum networks: A comprehensive survey},'' 2025-02-11 05:02:00 2025. [Online]. Available: \url{https://tsapps.nist.gov/publication/get_pdf.cfm?pub_id=958196}
\BIBentrySTDinterwordspacing

\bibitem{PERA}
\BIBentryALTinterwordspacing
H.~Hu, H.~Lun, Z.~Deng, J.~Tang, J.~Li, Y.~Cao, Y.~Wang, Y.~Liu, D.~Wu, H.~Yu, X.~Wang, J.~Wei, and L.~Shi, ``High-fidelity entanglement routing in quantum networks,'' \emph{Results in Physics}, vol.~60, p. 107682, 2024. [Online]. Available: \url{https://www.sciencedirect.com/science/article/pii/S2211379724003656}
\BIBentrySTDinterwordspacing

\bibitem{MULTIR}
Y.~Zeng, J.~Zhang, J.~Liu, Z.~Liu, and Y.~Yang, ``Entanglement routing design over quantum networks,'' \emph{IEEE/ACM Transactions on Networking}, vol.~32, no.~1, pp. 352--367, 2024.

\bibitem{Caleffi17}
M.~Caleffi, ``Optimal routing for quantum networks,'' \emph{IEEE Access}, vol.~5, pp. 22\,299--22\,312, 2017.

\bibitem{Pant20}
M.~Pant, H.~Krovi, D.~Towsley, L.~Tassiulas, L.~Jiang, P.~Basu, D.~Englund, and S.~Guha, ``Routing entanglement in the quantum internet,'' \emph{npj Quantum Information}, vol.~5, no.~1, Mar 2019.

\bibitem{Li21}
C.~Li, T.~Li, Y.-X. Liu, and P.~Cappellaro, ``Effective routing design for remote entanglement generation on quantum networks,'' \emph{npj Quantum Information}, vol.~7, no.~1, Jan 2021.

\bibitem{Chakraborty20}
K.~Chakraborty, D.~Elkouss, B.~Rijsman, and S.~Wehner, ``Entanglement distribution in a quantum network: A multicommodity flow-based approach,'' \emph{IEEE Transactions on Quantum Engineering}, vol.~1, pp. 1--21, 2020.

\bibitem{FMSPP}
Z.~Wang, J.~Li, K.~Xue, D.~S.~L. Wei, R.~Li, N.~Yu, Q.~Sun, and J.~Lu, ``An efficient scheduling scheme of swapping and purification operations for end-to-end entanglement distribution in quantum networks,'' \emph{IEEE Transactions on Network Science and Engineering}, vol.~11, no.~1, pp. 380--391, 2024.

\bibitem{PSC}
Z.~{Xiao}, J.~{Li}, K.~{Xue}, N.~{Yu}, R.~{Li}, Q.~{Sun}, and J.~{Lu}, ``{Purification scheduling control for throughput maximization in quantum networks},'' \emph{Communications Physics}, vol.~7, no.~1, p. 307, Dec. 2024.

\bibitem{PurSched}
\BIBentryALTinterwordspacing
Z.~Jia and L.~Chen, ``From entanglement purification scheduling to fidelity-constrained multi-flow routing,'' 2024. [Online]. Available: \url{https://arxiv.org/abs/2408.08243}
\BIBentrySTDinterwordspacing

\bibitem{Hu21}
\BIBentryALTinterwordspacing
X.-M. Hu, C.-X. Huang, Y.-B. Sheng, L.~Zhou, B.-H. Liu, Y.~Guo, C.~Zhang, W.-B. Xing, Y.-F. Huang, C.-F. Li, and G.-C. Guo, ``Long-distance entanglement purification for quantum communication,'' \emph{Phys. Rev. Lett.}, vol. 126, p. 010503, Jan 2021. [Online]. Available: \url{https://link.aps.org/doi/10.1103/PhysRevLett.126.010503}
\BIBentrySTDinterwordspacing

\bibitem{Pirandola17}
S.~Pirandola, R.~Laurenza, C.~Ottaviani, and L.~Banchi, ``{Fundamental limits of repeaterless quantum communications},'' \emph{Nature Commun.}, vol.~8, p. 15043, 2017.

\bibitem{Werner89}
\BIBentryALTinterwordspacing
R.~F. Werner, ``Quantum states with einstein-podolsky-rosen correlations admitting a hidden-variable model,'' \emph{Phys. Rev. A}, vol.~40, pp. 4277--4281, Oct 1989. [Online]. Available: \url{https://link.aps.org/doi/10.1103/PhysRevA.40.4277}
\BIBentrySTDinterwordspacing

\bibitem{Kalb17}
\BIBentryALTinterwordspacing
N.~Kalb, A.~A. Reiserer, P.~C. Humphreys, J.~J.~W. Bakermans, S.~J. Kamerling, N.~H. Nickerson, S.~C. Benjamin, D.~J. Twitchen, M.~Markham, and R.~Hanson, ``Entanglement distillation between solid-state quantum network nodes,'' \emph{Science}, vol. 356, no. 6341, pp. 928--932, 2017. [Online]. Available: \url{https://www.science.org/doi/abs/10.1126/science.aan0070}
\BIBentrySTDinterwordspacing

\bibitem{Dahlberg19}
\BIBentryALTinterwordspacing
A.~Dahlberg, M.~Skrzypczyk, T.~Coopmans, L.~Wubben, F.~Rozpundefineddek, M.~Pompili, A.~Stolk, P.~Pawe\l{}czak, R.~Knegjens, J.~de~Oliveira~Filho, R.~Hanson, and S.~Wehner, ``A link layer protocol for quantum networks,'' in \emph{Proceedings of the ACM Special Interest Group on Data Communication}, ser. SIGCOMM '19.\hskip 1em plus 0.5em minus 0.4em\relax New York, NY, USA: Association for Computing Machinery, 2019, p. 159–173. [Online]. Available: \url{https://doi.org/10.1145/3341302.3342070}
\BIBentrySTDinterwordspacing

\bibitem{Deutsch96Purification}
\BIBentryALTinterwordspacing
D.~Deutsch, A.~Ekert, R.~Jozsa, C.~Macchiavello, S.~Popescu, and A.~Sanpera, ``Quantum privacy amplification and the security of quantum cryptography over noisy channels,'' \emph{Phys. Rev. Lett.}, vol.~77, pp. 2818--2821, Sep 1996. [Online]. Available: \url{https://link.aps.org/doi/10.1103/PhysRevLett.77.2818}
\BIBentrySTDinterwordspacing

\bibitem{Waxman88}
B.~Waxman, ``Routing of multipoint connections,'' \emph{IEEE Journal on Selected Areas in Communications}, vol.~6, no.~9, pp. 1617--1622, 1988.

\bibitem{Coopmans21NetSquid}
T.~{Coopmans}, R.~{Knegjens}, A.~{Dahlberg}, D.~{Maier}, L.~{Nijsten}, J.~{de Oliveira Filho}, M.~{Papendrecht}, J.~{Rabbie}, F.~{Rozp{\k{e}}dek}, M.~{Skrzypczyk}, L.~{Wubben}, W.~{de Jong}, D.~{Podareanu}, A.~{Torres-Knoop}, D.~{Elkouss}, and S.~{Wehner}, ``{NetSquid, a NETwork Simulator for QUantum Information using Discrete events},'' \emph{Communications Physics}, vol.~4, no.~1, p. 164, Dec. 2021.

\bibitem{Wu21SeQUeNCe}
\BIBentryALTinterwordspacing
X.~Wu, A.~Kolar, J.~Chung, D.~Jin, T.~Zhong, R.~Kettimuthu, and M.~Suchara, ``Sequence: a customizable discrete-event simulator of quantum networks,'' \emph{Quantum Science and Technology}, vol.~6, no.~4, p. 045027, Sep. 2021. [Online]. Available: \url{http://dx.doi.org/10.1088/2058-9565/ac22f6}
\BIBentrySTDinterwordspacing

\end{thebibliography}

\end{document}